\documentclass[final,5p,times,twocolumn]{elsarticle}

\usepackage{graphicx}
\usepackage{amsmath}
\usepackage{amssymb}
\usepackage{xcolor}
\usepackage{soul}
\usepackage{float}
\usepackage{tikz}
\usetikzlibrary{arrows,matrix,positioning}

\usepackage[nottoc]{tocbibind}%Retirar na versão final

\usepackage{hyperref}
\hypersetup{
    colorlinks=true, %set true if you want colored links
    linktoc=all,     %set to all if you want both sections and subsections linked
    citecolor=blue,
    filecolor=black,
    linkcolor=blue,
    urlcolor=blue
}
\biboptions{comma,numbers,compress}

\journal{Materials Today Quantum}

\begin{document}

\begin{frontmatter}

%% Title, authors and addresses

%% use the tnoteref command within \title for footnotes;
%% use the tnotetext command for theassociated footnote;
%% use the fnref command within \author or \address for footnotes;
%% use the fntext command for theassociated footnote;
%% use the corref command within \author for corresponding author footnotes;
%% use the cortext command for theassociated footnote;
%% use the ead command for the email address,
%% and the form \ead[url] for the home page:
%% \title{Title\tnoteref{label1}}
%% \tnotetext[label1]{}
%% \author{Name\corref{cor1}\fnref{label2}}
%% \ead{email address}
%% \ead[url]{home page}
%% \fntext[label2]{}
%% \cortext[cor1]{}
%% \affiliation{organization={},
%%             addressline={},
%%             city={},
%%             postcode={},
%%             state={},
%%             country={}}
%% \fntext[label3]{}

\title{Exploring the Electronic Potential of Effective Tight$-$Binding Hamiltonians}

%% use optional labels to link authors explicitly to addresses:
%% \author[label1,label2]{}
%% \affiliation[label1]{organization={},
%%             addressline={},
%%             city={},
%%             postcode={},
%%             state={},
%%             country={}}
%%
%% \affiliation[label2]{organization={},
%%             addressline={},
%%             city={},
%%             postcode={},
%%             state={},
%%             country={}}

\author[inst1]{Grazi\^{a}ni Candiotto}

\affiliation[inst1]{organization={Instituto de Física, Universidade Federal do Rio de Janeiro},%Department and Organization
           % addressline={Address One}, 
            city={Rio de Janeiro},
            postcode={21941-972}, 
            state={RJ},
            country={Brazil}}

\begin{abstract}
%% Text of abstract
The linear combination of atomic orbitals (LCAO) is a standard method for studying solids and molecules, it is also known as the tight$-$binding (TB) method. In most of the implementations only the basis set and the coupling constants are provided, without the explicit definition of kinetic and potential energy operators. The tight$-$binding scheme is, nonetheless, capable of providing an accurate description of properties such as the electronic bands and elastic constants for many materials. However, for some applications, the knowledge of the underlying electronic potential associated with the tight$-$binding hamiltonian might be important to guarantee that the actual physics is preserved by the semiempirical scheme. In this work the electronic potentials that arise from the use of tight$-$binding effective hamiltonians it is explored. The formalism is applied to the extended H\"{u}ckel tight$-$binding (EHTB) hamiltonian, which is a two$-$center Slater$-$Koster approach that makes explicit use of the overlap matrix.
\end{abstract}

%%Graphical abstract
%\begin{graphicalabstract}
%\includegraphics{grabs}
%\end{graphicalabstract}

%%Research highlights
%\begin{highlights}
%\item Research highlight 1
%\item Research highlight 2
%\end{highlights}

\begin{keyword}
%% keywords here, in the form: keyword \sep keyword
Effective potential \sep Tight$-$binding
%% PACS codes here, in the form: \PACS code \sep code
%\PACS 0000 \sep 1111
%% MSC codes here, in the form: \MSC code \sep code
%% or \MSC[2008] code \sep code (2000 is the default)
%\MSC 0000 \sep 1111
\end{keyword}

\end{frontmatter}

%% \linenumbers

%% main text
%-------------------------------------------------------------------------------
\section{INTRODUCTION}\label{sec:introduction}
%-------------------------------------------------------------------------------

The linear combination of atomic orbitals (LCAO), also known as tight$-$binding (TB) method, is largely used for describing the transport properties in condensed matter,\cite{fu2007topological,eschrig2009tight,renani2011ligand,kienle2006extended,mireles2001ballistic} for studying the electronic structure of complex solids\cite{cohen1994tight,cerda2000accurate,mehl1996applications} and large molecular systems,\cite{amara2009tight,renani2012tight,monti2015crucial} for performing molecular dynamics,\cite{medrano2016photoinduced} among other applications. Several TB prescriptions have been developed since the original proposal of Slater and Koster.\cite{slater1954simplified} Noteworthy, the tight$-$binding density functional theory (TB$-$DFT) method has been used as an efficient alternative for the calculation of the electronic structure and for molecular dynamics simulations of large systems.\cite{foulkes1989tight,niehaus2001tight,porezag1995construction,pal2016nonadiabatic} The TB or LCAO schemes consist of making a linear combination of atomic orbitals, localized on the atomic positions, as a means for developing the bonding interactions in terms of site$-$centered orbitals. Due to its simplicity and versatility, much knowledge has been gained through the use of tight$-$binding methods. For instance, the method is capable of describing accurately the total energy of solids and molecules, the energy differences caused by small structure deformations and the associated phonon spectra.\cite{cerda2000accurate} It can also reproduced the equations of state and defect energies,\cite{mehl1996applications,cohen1997tight} among other structural properties.

To best describe the properties of the system a set of TB parameters must be tuned so that the calculations fit some reference data set, which may either be obtained from higher level theory or gathered from experiments. If well parametrized, the method can accurately reproduce \textit{ab$-$initio} calculations for a variety of materials $-$ including transition and noble metals,\cite{cohen1997tight,tang1996environment} semiconductors,\cite{cerda2000accurate} and molecules \cite{renani2011ligand} $-$ but with a much lower computational cost. Therefore complex problems and materials can be studied for the first time.

However, for some applications, it might be necessary to discern the actual couplings that are implied by the tight$-$binding parameters, because the effective TB hamiltonian aims at describing only the total energy of the state, disregarding whether it is kinetic or due to a potential. The knowledge of the underlying electronic potential associated with the tight$-$binding method is important to guarantee that the actual physics is preserved by the semiempirical scheme, since several parameter sets can frequently fit the data. In this work the electronic potentials that result from the use of tight$-$binding effective hamiltonians is explored. The formalism is applied to the extended H\"{u}ckel tight$-$binding (EHTB) hamiltonian, which has been widely used by physicists, chemists and materials scientist to describe the electronic structure and properties of molecules and solids. The EHTB method corresponds to a two$-$center Slater$-$Koster approach that makes explicit use of the overlap matrix, which brings to the model information about disorder, molecular shape, molecular arrangement, as well introduce interactions between atomic orbitals in a molecule or solid.

The EHTB method is in essence a two$-$center Slater$-$Koster formulation of tight$-$binding with a nonorthogonal basis. According to the extended H\"{u}ckel prescription, the elements of the hamiltonian matrix are written as a linear function of the overlap (or adjacency) matrix \textbf{S}. Thus, for a given basis set

\begin{equation}
\begin{split}
\left\langle i \vert \mathbf{H} \vert j \right\rangle = \xi_{ij}\left\langle i\vert j\right\rangle,\\
H_{ij}=\xi_{ij}S_{ij},
\end{split}
\end{equation}

\noindent where $\left\lbrace \vert i\rangle \right\rbrace$ is some generic local basis set $\chi_{i}(\mathbf{r})=\left\langle\mathbf{r}\vert i\right\rangle$. $\xi_{ij}$ is commonly written as

\begin{equation}
\xi_{ij} = K_{ij}\left(\dfrac{E_{i}+E_{j}}{2}\right),
\end{equation}

\noindent where $K_{ij}$ is a coupling parameter, which is usually set to 1.75 if $i\neq j$, or 1 if $i=j$. The diagonal elements of $\mathbf{H}$, the $\mathbf{E}_{i/j}$ parameters, may be chosen to represent the valence ionization potentials of the atomic species, or the on$-$site energies associated with the localized basis set. That way a small set of semi$-$empirical parameters suffice to build up the hamiltonian, which renders the EHTB method a clear physical interpretation, with a reduced number of semiempirical parameters (as compared to the SK method) and good transferability, because it makes explicitly use of the overlap matrix. Thereby, the method accounts approximately for effects produced by geometrical and chemical bond variations.\cite{monti2015crucial} The EH hamiltonian can be mapped onto the traditional TB Hamiltonian form,

\begin{equation}
\widehat{H}_{tb} = \sum_{i} \epsilon_{i}\widehat{a}_{i}^{\dagger}\widehat{a}_{i} - \sum_{i\neq j}t_{ij}\widehat{a}_{i}^{\dagger}\widehat{a}_{j} + c.c.,
\end{equation}

\noindent by choosing the parameters

\begin{equation}
\begin{split}
\epsilon_{i} &= K_{ii}E_{i},\\
-t_{ii} &= K_{ij}\left(\dfrac{E_{i}+E_{j}}{2}\right)S_{ij}.
\end{split}
\end{equation}

\noindent The diagonal elements of $\mathbf{H}$ are called Coulomb Integrals; they represent the combined kinetic and potential energies of an electron described by the orbital $\chi_{i}(\mathbf{r})$, experiencing the electrostatic interactions with all the other electrons and all the positive nuclei. The off$-$diagonal elements are called Resonance (or Bond) Integrals.\cite{hoffmann2004theory} In addition to the Slater$-$type orbitals originally proposed by Hoffmann,\cite{hoffmann2004theory} other simple algebraic forms may be used for basis set: for instance, Gaussian or harmonic oscillator basis functions. The EHT, when compared to the traditional SK tight$-$binding method, provides good accuracy while presenting several important advantages:\cite{kienle2006extended,cerda2000accurate} i) a considerable reduction in the number of parameters to be fitted, ii) natural scaling laws for the coupling parameters and, iii) an improved transferability of the parameters.

In the remainder of the paper it is described, in section \ref{sec:ETBP}, the formalism used to obtain the effective electronic potential associated with the Extended H\"{u}ckel tight$-$binding (EHTB) hamiltonian is presented and applied in the calculation it for several prototypical molecular systems. In section \ref{sec:QDA} the method is applied to a model for two$-$dimensional quantum dot arrays and in section \ref{sec:conclusions} the main conclusions of this study are presented.

%--------------------------------------------------------------------------------
\section{Electronic Tight$-$Binding Potential}\label{sec:ETBP}
%--------------------------------------------------------------------------------

The tight$-$binding hamiltonian is not constructed from the kinetic $(\widehat{T})$ or Coulomb potential $(\widehat{V})$ operators but it is, otherwise, directly determined by the basis set functions $\chi_{i}(\mathbf{r})$, or simply built up from a set of appropriate coupling parameters. As a result, several parameter sets can frequently fit the reference set, with different electronic potentials associated to them. It is, therefore, desirable to know that the underlying physics is preserved by the parametrization scheme. In this section it is presented a procedure to determine the effective electronic potential that results from the use of a tight$-$binding hamiltonian.

For an operator $\widehat{O}$, written in terms of a nonorthogonal basis set, one has

\begin{equation}
\begin{split}
O(\mathbf{r},\mathbf{r^{\prime}}) &=\left\langle \mathbf{r}\left\vert \hat{O} \right\vert \mathbf{r^{\prime}}\right\rangle,\\
&=\langle \mathbf{r}\vert \left[\sum_{n,i}\vert n\rangle S_{ni}^{-1}\langle i \vert\right] \widehat{O}\left[ \sum_{j,m}\vert j\rangle S_{jm}^{-1}\langle m \vert \right]\vert \mathbf{r^{\prime}}\rangle,\\
&=\sum_{n,m \atop i,j}\langle \mathbf{r}\vert n\rangle S_{ni}^{-1}\langle i \vert \widehat{O} \vert j\rangle S_{jm}^{-1}\langle m \vert \mathbf{r^{\prime}}\rangle,\\
&=\sum_{n,m \atop i,j}\chi_{n}(\mathbf{r})\, S_{ni}^{-1}\, O_{ij}\, S_{jm}^{-1}\,\chi_{m}(\mathbf{r^{\prime}}), 
\end{split}
\end{equation}

\noindent where the sums are carried over all basis states. By defining $\mathbf{\widetilde{O}= S^{-1}OS^{-1}}$ it is avoided the trouble of dealing with the nonorthogonality of the basis, thus

\begin{equation}
O(\mathbf{r},\mathbf{r^{\prime}})=\sum_{n,m}\chi_{n}(\mathbf{r})\, \widetilde{O}_{nm}\, \chi^{*}_{m}(\mathbf{r^{\prime}}).
\end{equation}

\noindent The operator $O(\mathbf{r},\mathbf{r^{\prime}})$ acounts for the interference of wavefunctions $\chi_{n}(\mathbf{r})$ and $\chi^{*}_{m}(\mathbf{r^{\prime}})$. The operator $\widehat{O}$ can be formally written in the $\vert n \rangle$ representation as

\begin{equation}
\widehat{O} = \sum_{n,m} \vert n\rangle\, \widetilde{O}\, \langle m\vert,
\end{equation}

\noindent as well as

\begin{equation}
\widehat{O} = \int\int \vert \mathbf{r}\rangle O(\mathbf{r},\mathbf{r^{\prime}}) \langle \mathbf{r^{\prime}}\vert\: d\mathbf{r}\: d\mathbf{r^{\prime}},
\end{equation}

\noindent in the representation of the coordinates. However, if the single particle operator $\widehat{O}$ happens to be a local function of $\widehat{r}$, \textit{i.e.}, $\widehat{O} = f(\widehat{r})$ (like the Coulomb or the harmonic oscillator potentials), then $\widehat{O}$ must be diagonal in the coordinates

\begin{equation}
\begin{split}
\widehat{O} &=\int\int \vert\mathbf{r} \rangle\, O(\mathbf{r},\mathbf{r^{\prime}})\, \delta(\mathbf{r}-\mathbf{r^{\prime}})\, \langle\mathbf{r^{\prime}} \vert \, d\mathbf{r}\,d\mathbf{r^{\prime}},\\
&=\int \vert\mathbf{r} \rangle\, f(\mathbf{r})\, \langle\mathbf{r} \vert \, d\mathbf{r}.
\end{split}
\end{equation}

\noindent So far it has been assumed a complete basis set $\vert n \rangle$, but if this is not the case, it is possible to write

\begin{equation}
\begin{split}
\widehat{O}\approx \widehat{O}_{N} &\equiv\sum^{N}_{n,m}\vert n\rangle\, \widetilde{O}_{nm}\, \langle m\vert,\\
&\equiv\sum^{N}_{k}\widehat{O}^{(k)},
\end{split}
\end{equation}

\noindent where $N$ is the truncation order and

\begin{equation}\label{eq:truncation_order}
\begin{split}
\widehat{O}^{(k)} &=\sum^{k}_{n,m}\vert n\rangle\, \widetilde{O}_{nm}\left(\dfrac{\delta_{nk}+\delta_{mk}}{1+\delta_{nk}\delta_{mk}}\right)\, \langle m\vert,\\
&=\sum^{k}_{n,m}\vert n\rangle\, \widetilde{O}_{nm}\Delta^{(k)}_{nm}\, \langle m\vert,
\end{split}
\end{equation}

\noindent as depicted below

\begin{equation}
\widehat{O}= \begin{pmatrix} \begin{tikzpicture}
\matrix [matrix of math nodes] (m)
{
    \widetilde{O}_{11} & \widetilde{O}_{12} & \widetilde{O}_{13} & \cdots \\
    \widetilde{O}_{21} & \widetilde{O}_{22} & \widetilde{O}_{23} & \cdots \\
    \widetilde{O}_{31} & \widetilde{O}_{32} & \widetilde{O}_{33} & \cdots \\     
          \vdots       &      \vdots        &       \vdots       & \ddots  \\  
};  
        \draw[color=black] (m-1-1.north west) -- (m-1-3.north east) -- (m-3-3.south east) -- (m-3-1.south west) -- (m-1-1.north west);
        \draw[color=black] (m-1-1.north west) -- (m-1-2.north east) -- (m-2-2.south east) -- (m-2-1.south west) -- (m-1-1.north west);
        \draw[color=black] (m-1-1.north west) -- (m-1-1.north east) -- (m-1-1.south east) -- (m-1-1.south west) -- (m-1-1.north west);
    \end{tikzpicture}
    \end{pmatrix}.
\end{equation}

\noindent In Eq. (\ref{eq:truncation_order}) the numerator selects the $k-$th column, and the denominator normalizes the factor 2 that appears for diagonal elements, $i = j = k$. It will be shown in section \ref{sec:QDA} that the truncation of this operator affects the nonlocality of the potential. By using the TB hamiltonian one has no knowledge of the kinetic and potential energy components of the particles, except for the fact that, in principle, the eigenstates diagonalize some effective hamiltonian $H_{\chi} = T_{\chi} + V_{\chi}$. Only $H$ is known in the representation of the truncated basis set functions $\lbrace \chi_{i}(\mathbf{r})\rbrace$. However, it is possible to apply the Kohn$-$Sham ansatz \citep{kohn1965self} and assume that the kinetic energy is equal to the kinetic energy of a system of a non$-$interacting particles. By writing the kinetic energy (in atomic units) as

\begin{equation}
T_{ij}=\langle i \vert \nabla^{2} \vert j\rangle,
\end{equation}

\noindent so it is possible to introduce an effective electric potential $\mathbf{V}$, which is defined as

\begin{equation}
\begin{split}
V_{ij} &= H_{ij} - T_{ij},\\
&=\xi_{ij}S_{ij}-T_{ij}.
\end{split}
\end{equation}

\noindent Thus, it becomes possible to map the TB hamiltonian onto a dynamic system, where it is possible to calculate the kinetic energy and the potential $\mathbf{V}$.

Thus far, $\mathbf{T}$ and $\mathbf{V}$ were determined in the representation of basis $\lbrace\vert n\rangle\rbrace$, however $V(\mathbf{r})$ is more amenable to interpretation than $V_{ij}$. Then, rewrite the potential in the representation of the coordinates as

\begin{equation}\label{eq:Vseries}
V(\mathbf{r},\mathbf{r^{\prime}})=\sum_{n,m}\chi_{n}(\mathbf{r})\widetilde{V}_{nm}\chi^{*}_{m}(\mathbf{r^{\prime}}).
\end{equation}

Although $V(\mathbf{r},\mathbf{r^{\prime}})$ should be local for an independent particle, that is not the case due to the approximations undertaken, which are the basis space truncation and the use of the KS ansatz. Though it is not possible to assume that $V(\mathbf{r},\mathbf{r^{\prime}})$ is diagonal, one can expect that this is a reasonable approximation, since the local density approximation (LDA)\cite{kohn1965self,perdew1981} is known to work well for systems without strong correlations. Therefore it is defined

\begin{equation}\label{eq:veff}
V_{eff}(\mathbf{r})=V(\mathbf{r},\mathbf{r}).
\end{equation}

\noindent This local form will be used in the following to obtain the effective potential of the EHTB method.

\begin{figure}[t!]
    \centering
    \includegraphics[width=\linewidth]{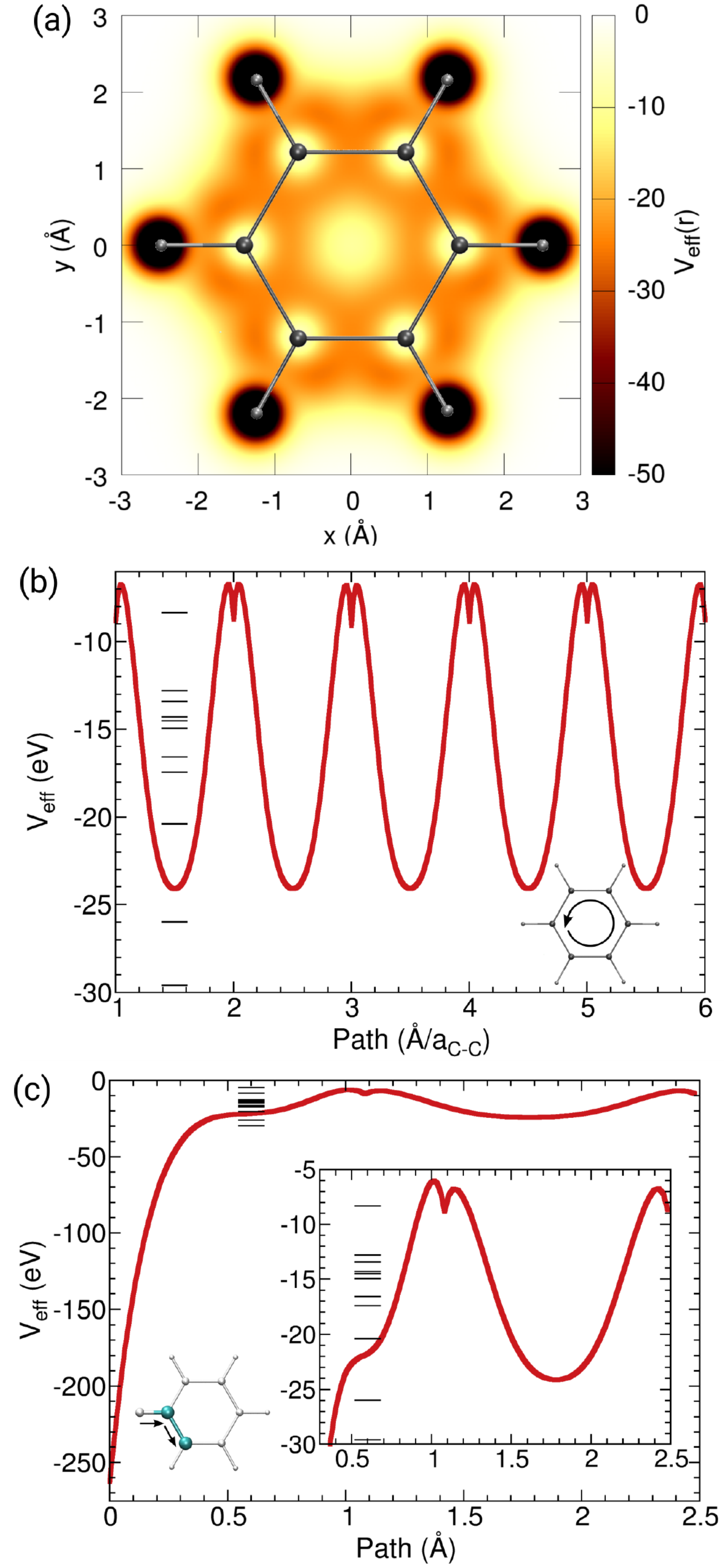}
\caption{EHT electronic potential $V_{eff}(\mathbf{r})$ for the benzene molecule: a) surface map of $V_{eff}(\mathbf{r})$; b) $V_{eff}(\mathbf{r})$ along the contour of carbon cycle, the distance along the path is measured in units of the C$-$C. The horizontal lines show the energies of the molecular orbitals (eigenstates of the EHT hamiltonian) up to the HOMO level; c) the potential along the H$-$C$-$C path, with distance measured in \AA.}\label{fig:1}
\end{figure}

%--------------------------------------------------------------------------------
\subsection{The Electronic Potential of the Extended H\"{u}ckel Tight$-$Binding Method}
%--------------------------------------------------------------------------------

Having defined the effective electronic potential operator, the EHTB $V_{eff}(\mathbf{r})$ was calculated for some prototypical molecules. Recalling that the functional form of the electronic potential shall be determined by both the tight$-$binding parameters as well as the basis set. The basis set functions generally employed to describe chemical species are the Slater$-$type orbitals (STO), defined as

\begin{equation}\label{eq:stos_wavefunction}
\chi_{nml}(\mathbf{r})=r^{(n-1)}e^{-\zeta r}Y^{m}_{l}\left(\theta,\phi\right).
\end{equation}

\noindent For the STO’s, $\zeta$ is a semiempirical parameter related to the effective charge of the ion, $n$ is the main quantum number and the $Y^{m}_{l}\left(\theta,\phi\right)$ are spherical harmonics. The radial part of the wavefunctions (Eq. \ref{eq:stos_wavefunction}) characterize the Slater$-$type orbitals. To simplify the notation in the following equations, the indices $n$, $l$, $m$, and $\zeta$, are combined into one global index $N$, and write $\vert N\rangle = \vert n l m;\zeta\rangle$.

The kinetic energy matrix elements between two STOs centered at $\mathbf{R}_{A}$ and $\mathbf{R}_{B}$ can be written as a sum of overlap elements (see ref. \cite{rico1989molecular}),

\begin{equation}
\begin{split}
T^{N}_{N^{\prime}} &=-\dfrac{1}{2}\langle n^{\prime} l^{\prime} m^{\prime};\zeta^{\prime} \vert \nabla^{2} \vert n l m;\zeta \rangle,\\
&=-\dfrac{1}{2}\left\lbrace \left[n\left(n-1\right)-l\left(l+1\right)\right]S^{\left(n-2\right)lm}_{n^{\prime} l^{\prime} m^{\prime}} \right. \\
 &\left.\qquad\qquad\qquad\qquad-2 n \zeta S^{\left(n-1\right)lm}_{n^{\prime} l^{\prime} m^{\prime}}+\zeta^{2}S^{N}_{N^{\prime}} \right\rbrace.
\end{split}
\end{equation}

In the calculations that follow, standard Hoffmann parameters are used for the H, C, N, O and S atoms,\cite{alvarez2012tables} unless explicitly noted otherwise. The basis set includes the $3s$ and $3p$ atomic orbitals for S atoms, $2s$ and $2p$ atomic orbitals for C, N and O atoms, and the $1s$ atomic orbital for H atoms. 

Starting by analysing the Extended H\"{u}ckel electronic potential of the benzene molecule, as presented in Figure \ref{fig:1}, where the top image shows a surface map of $V_{eff}(\mathbf{r})$; graphs b) and c) show the EHTB potential along different paths in the molecular structure, as indicated by the insets. This figure evinces the general features of the EHTB potential produced by the Slater$-$type orbitals. Along the carbon ring (Figure \ref{fig:1}b), the potential minima is located in the bond region $-$ not on the site of the C atoms $-$ because only the valence states ($2s$ and $2p$) are used to describe these atoms. The former orbitals are mostly involved with the formation of bonds, via the resonance (off$-$diagonal) terms of the extended H\"{u}ckel hamiltonian. The effect it produces on the EHTB potential is analogous to the effect produced by the use of pseudo$-$potentials in first$-$principles calculations, \cite{pickett1989pseudopotential} which are employed to avoid the ionic potential at the core region. The potential is a lot deepper, however, on the site of the H atom (Figure \ref{fig:1}c), since it is the result of only the $1s$ orbital. The energies of the molecular orbitals (MO), up to the HOMO level, as obtained by solving the EHTB eigenvalue equation for the C$_{6}$H$_{6}$ molecule, are depicted by horizontal bars alongside $V_{eff}(\mathbf{r})$. Despite the deep quantum wells on the hydrogen sites the energies of the MOs are comparable to the bonding potentials of the carbon ring.

\begin{figure}[b!]
    \centering
    \includegraphics[width=\linewidth]{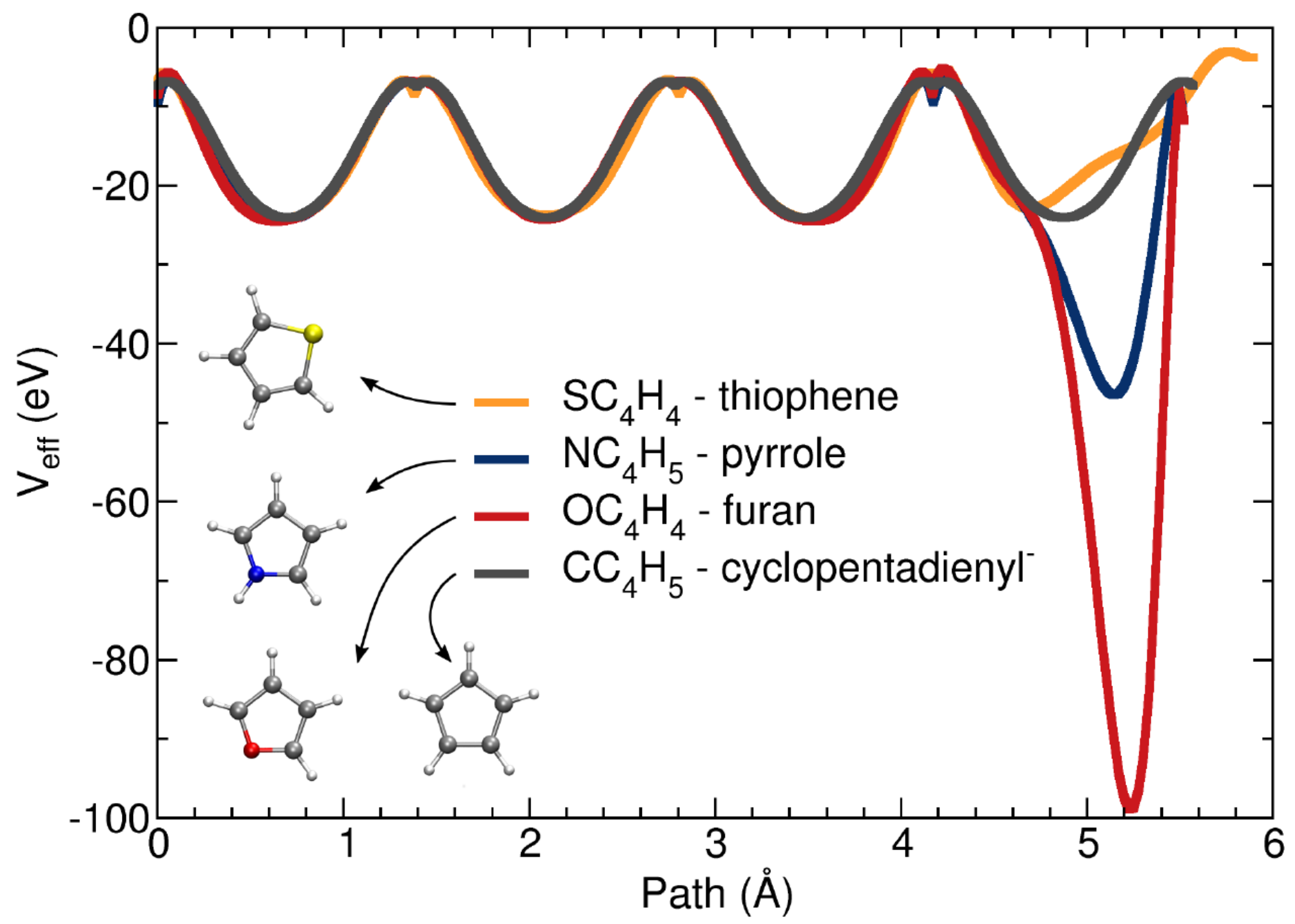}
\caption{EHT electronic potential $V_{eff}(\mathbf{r})$ for the pentagon$-$shaped molecules: thiophene (SC$_{4}$H$_{4}$, orange), pyrrole (NC$_{4}$H$_{5}$, blue), furan (OC$_{4}$H$_{4}$, red) and cyclopentadienyl (C$_{5}$H$_{5}$, black). $V_{eff}(\mathbf{r})$ along the contour of the ring, with distance measured in \AA.}\label{fig:2}
\end{figure}

To analyze the effect produced by different atoms it was calculated $V_{eff}(\mathbf{r})$ for four pentagonal molecules of distinct compositions, namely: thiophene (SC$_{4}$H$_{4}$), pyrrole (NC$_{4}$H$_{5}$), furan (OC$_{4}$H$_{4}$) and cyclopentadienyl (C$_{5}$H$_{5}$). Their EHTB potentials are shown in Figure \ref{fig:2}, color coded as: SC$_{4}$H$_{4}$ (orange), NC$_{4}$ H$_{5}$ (blue), OC$_{4}$H$_{4}$ (red) and C$_{5}$H$_{5}$ (black). Along the molecular ring the $V_{eff}(\mathbf{r})$ shows the same characteristics as in the benzene molecule (Figure \ref{fig:1}), except on the positon of the heteroatom. In this case, the effective potential is deeper for the C$-$O bond, followed by C$-$N, C$-$C and C$-$S, reflecting the different electronegativity of the elements ($3.44$ for O, $3.04$ for N, $2.55$ for C and $2.58$ for S).

\begin{figure}[t!]
    \centering
    \includegraphics[width=\linewidth]{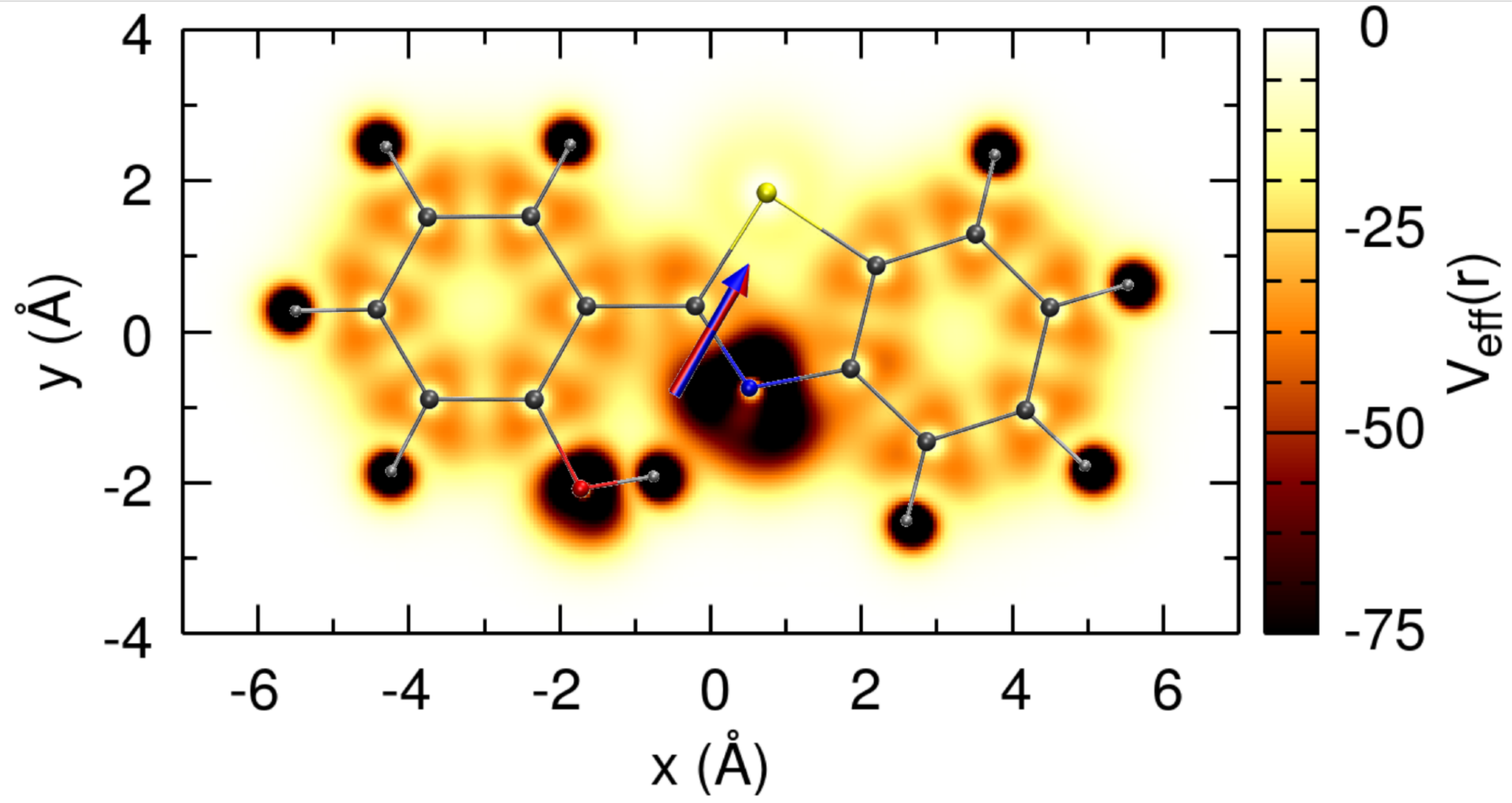}
\caption{Surface map of the EHT electronic potential $V_{eff}(\mathbf{r})$ for the 2$-$(2’$-$hydroxyphenyl)benzothiazole (HBT) molecule (geometry superposed). The arrows describe the electric dipole moments, as calculated with DFT ($1.99$ Debye, red) and EHT ($2.02$ Debye, blue).}\label{fig:3}
\end{figure}

A more demanding calculation was performed for the 2$-$(2’$-$hydroxyphenyl)benzothiazole (HBT) molecule, shown in Figure \ref{fig:3}. In this case, the standard EHTB parameters did not yield a good description of the molecule’s HOMO$-$LUMO gap, neither of its dipole moment. Thus, in order to obtain parameters that are more consistent with the HBT molecule, a functional density theory (DFT) was used to obtain the dipole moment and the HOMO$-$LUMO energy gap as references. First the HBT molecule was relaxed to ground state geometry using a genetic$-$algorithm scheme\cite{haupt2004practical}. In DFT calculations, the hybrid functional Becke three$-$parameter Lee$-$Yang$-$Parr (B3LYP)\cite{becke93,candiotto2020} was used along with a split$-$valence double$-$zeta polarized Gaussian type orbital basis, 6$-$31G ($d$, $p$)\cite{ditchfield2003}. For the sake of comparison with the \textit{ab$-$initio} theory it is presented the dipole moments, $1.99$ Debye and $2.02$ Debye for the DFT and EHT methods, respectively, and the HOMO$-$LUMO gap of $4.14$ eV for DFT and $4.22$ eV for EHT. Figure \ref{fig:3} presents the surface map of $V_{eff}(\mathbf{r})$, with two arrows superposed to it, which represent the dipole moment vectors obtained by the DFT (red) and EHT (blue) calculations. Both the orientation and size of the dipole moments show excellent agreement. It is interesting to notice also that the effective potential exhibits a strong gradient along the same direction as the electronic dipole moments, that is, deeper at the OH enol group and shallower on the S atom of the thiazole heterocycle. That potential gradient gives rise to electron concentration on the OH enol group and charge depletion around the S atom.

\begin{figure}[b!]
    \centering
    \includegraphics[width=\linewidth]{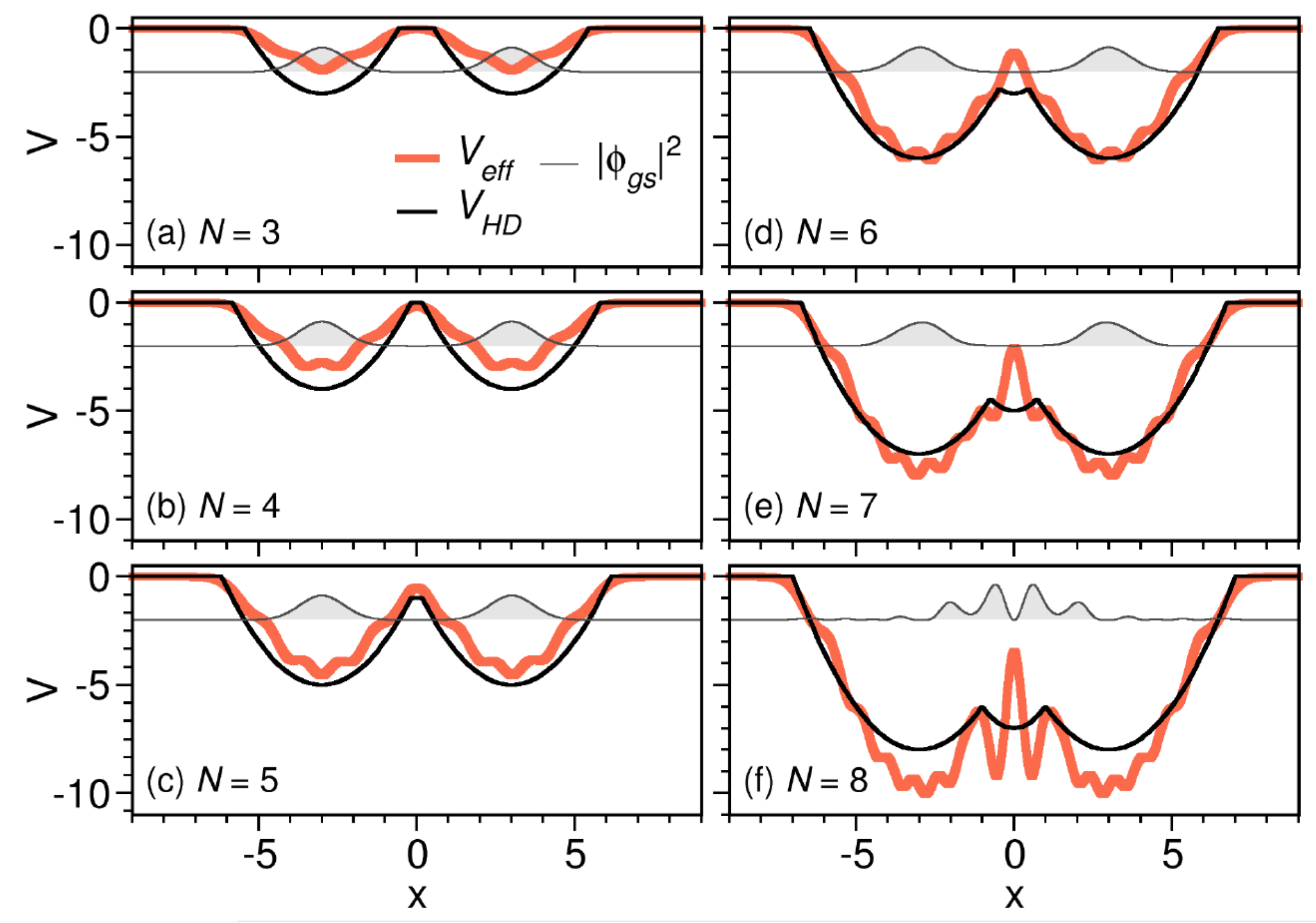}
\caption{Effective potential $V_{eff}(x)$, harmonic dimer potential $V_{HD}(x)$ and ground state probability density $\vert\phi_{gs}(x)\vert^{2}$ (grey shade), for various basis set sizes: from $N = 3$ to $N = 8$, in (a) to (f), respectively. ($\vert\phi_{gs}(x)\vert^{2}$ is scaled by two and arbitrarily shifted for easier visualization).}\label{fig:4}
\end{figure}

%--------------------------------------------------------------------------------
\section{QUANTUM DOT ARRAYS}\label{sec:QDA}
%--------------------------------------------------------------------------------

The linear combination of harmonic functions associated to localized parabolic potentials can describe charge and energy transport in quantum dot arrays, \cite{nazir2005anticrossing,kamat2008quantum} artificial molecules and mesoscopic devices, \cite{popsueva2007structure,kagan2015charge,kagan2015charge} organic and supramolecular structures,\cite{adronov2000light,teyssandier2016Host,scholes2011lessons} and photonic crystals \cite{hashemi2010emergence}. In this section it is applied the effective potential method to a system comprised of localized parabolic quantum wells.

Firstly it is analysed a model of two coupled parabolic wells, as shown in Figure \ref{fig:4}, with the goal of comparing the obtained  $V_{eff}(\mathbf{r})$ with the actual potential profile. Assuming that the electronic structure of the harmonic dimer is given by the EHTB model, calculated on the basis of the harmonic oscillator eigenstates

\begin{equation}\label{eq:gaussian_wavefunction}
\chi_{n}(x)=\dfrac{1}{\sqrt{2^{n}n!}}\left(\dfrac{m\omega}{\pi}\right)^{1/4}\exp\left(-\dfrac{m\omega x^2}{2\hbar}\right)H_{n}\left(\alpha x\right),
\end{equation}

\noindent where $H_{n}$ is the Hermite polynomial of argument $\alpha = \sqrt{m\omega/\hbar}$ and $n$ order, associated with the energy eigenvalue

\begin{equation}
E_{n}=\hbar\omega\left(n+\dfrac{1}{2}\right).
\end{equation}

\noindent To assure that the energy of the bound states are negative, it is shifted the Extended H\"{u}ckel hamiltonian by the value $E_{shift}=\hbar \omega(N+1/2)$, as follows

\begin{equation}\label{eq:hamiltonian_shifited}
H_{ij}=\left[K_{ij}\left(\dfrac{E_{i}+E_{j}}{2}\right)-E_{shift}\right]S_{ij},
\end{equation}

\noindent where the integer $N$ designates the number of bound states. The elements $S_{ij}$ of the overlap matrix can be calculated analitically for the orbitals (Eq. \ref{eq:gaussian_wavefunction}). In Figure \ref{fig:4} we show the effective potential $V_{eff}$ (red curve) and the actual potential of the harmonic dimer (HD) $V_{HD}$ (black line), for an increasing number of bound states: from $N = 3$ to $N = 8$. Notice that the intersection of the two parabolic wells gives rise to another parabolic well centered at the origin. We observe that $V_{eff}$ and $V_{HD}$ differ more significantly if the number of basis states is too few ($N < 5$) or too many ($N > 7$). In the former case there are not enough basis functions to describe the bound states whereas, in the later, the delocalized basis functions start to interfere. The probability density of the ground ($\vert\phi_{gs}(x)\vert^{2}$) is also shown, scaled by two and arbitrarily shifted for the sake of visualization. These eigenstates are solutions of the EHTB eigenvalue equation $\mathbf{HC} = E\mathbf{SC}$, obtained for the hamiltonian (Eq. \ref{eq:hamiltonian_shifited}) in the nonorthogonal basis set $\lbrace\chi_{n}\rbrace$, where $\mathbf{C}$ is the vector associated to $\phi(x)=\sum_{n}C_{n}\chi_{n}(x)$.

\begin{figure}[t!]
    \centering
    \includegraphics[width=\linewidth]{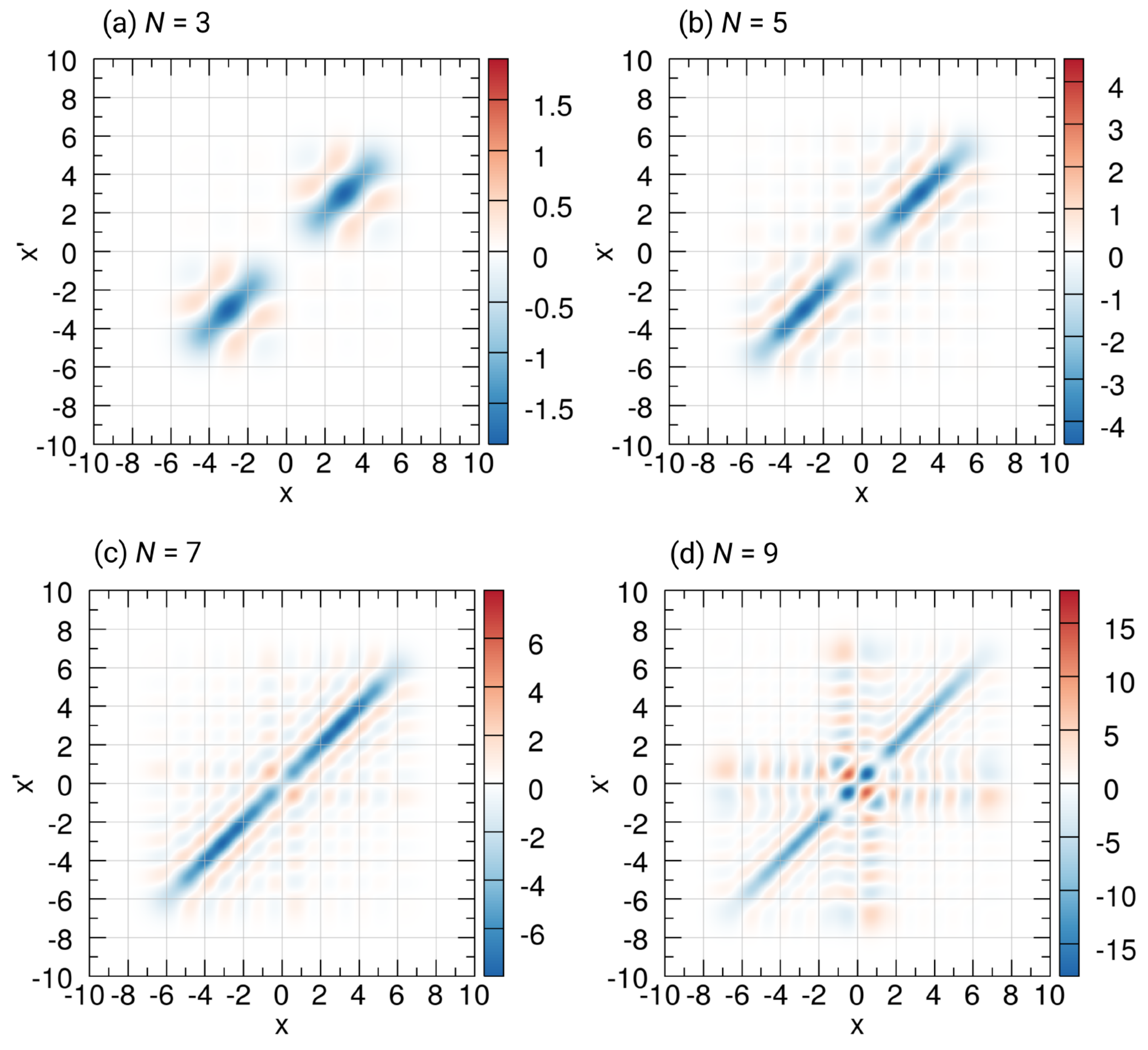}
\caption{Potential operator in coordinates representation, $V(x,x^{\prime})$ for $N = 3, 5, 7, 9$.}\label{fig:5}
\end{figure}

Although $V_{eff}(\mathbf{r})$ is defined as a local potential by Eq. \ref{eq:veff}, the effective potential is actually nonlocal, as given by Eq. \ref{eq:Vseries}. For the nonlocal potential, the energy term $U(x)=V(x)\phi(x)$ in the Schr\"{o}dinger equatoin is replaced by

\begin{equation}
U(x)=\int V(x,x^{\prime})\psi(x^{\prime})dx^{\prime}
\end{equation}

\noindent where the kernel $V(x,x^{\prime})$ contains the nonlocal dependence. So far the local (diagonal) part of the effective potential has been shown, but in Figure \ref{fig:5} it is showed the behavior of the nonlocal effective potential for the harmonic dimer, for basis sizes $N =3, 5, 7$ and $9$. From the physical viewpoint, nonlocal potentials can alter the solution of a local double$-$well potential by producing wavefunctions with more nodes than those of the local potential,\cite{hooverman1972anomalous} by changing the size of the local barrier and, consequently, the tunneling dynamics.\cite{razavy2013quantum} These effects can give rise to anomalous diffusion\cite{lenzi2008solutions} and, therefore, affect the calculated transport properties of the system. For the harmonic dimer presently considered (Figure \ref{fig:5}) the nonlocality is of short range and does not couple the two wells until they start to merge,for $N\geq 8$, but if the coupling parameter increses the nonlocality of the effective potential is expected to increase.

\begin{figure}[t!]
    \centering
    \includegraphics[width=\linewidth]{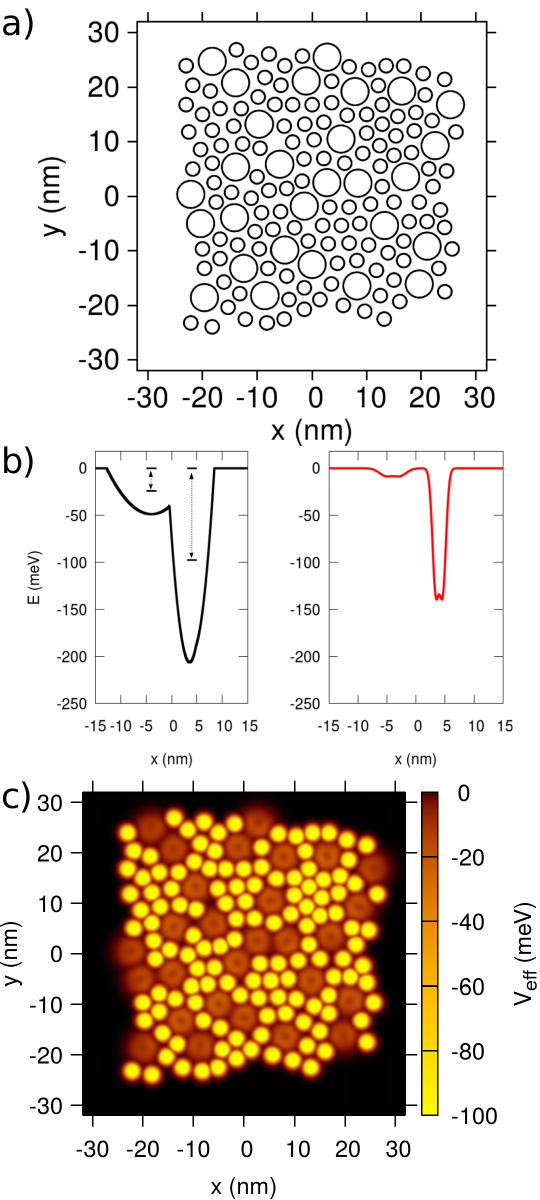}
\caption{a) Two$-$dimensional array of quantum dots, where the large dots have a radius $R_{1} = 2.5$ nm and the small ones $R_{2} = 1.25$ nm. b) Left side, black curve: parabolic potentials and the characteristic energies $\omega_{1}$ and $\omega_{2}$. Right side, red curve: the EHTB effective electronic potential, $V_{eff}(r)$. c) The EHTB effective electronic potential  $V_{eff}(r)$ for the 2D$-$QD array.}\label{fig:6}
\end{figure}

At this moment the EHTB method it will be applied to calculate the effective potential for a two$-$dimensional array of quantum dots (2D$-$QD). For this application, a natural photosynthetic complex present in photosynthetic membrane of \textit{Rhodospirillum photometricum}) was chosen. The natural photosynthetic complex used in the calculations was taken from an atomic force microscopy (AFM) image present in the ref. \cite{scholes2011lessons}. This supramolecular structure containing around 160 subunits. This system is basically characterized by two distinct subunits called light harvest complex I (LHC$_{I}$, large rings) and light harvest complex II (LHC$_{II}$, small rings). The subunit LHC$_{II}$ is responsible for collecting (most of) the light that reaches the membrane, converting it into excitonic energy and transferring this energy to the LHC$_{I}$ subunit where the reaction center is located, which is responsible for converting excitonic energy into chemical energy. On the other hand, the LHC$_{I}$ subunit is responsible for collecting (part of) the light that reaches until the membrane, converting it into excitonic energy and transferring it to the reaction center, as well as receiving the excitonic energy originating from the LHC$_{II}$ subunits. Thus, in the model each subunit (LHC$_{I}$ and LCH$_{II}$) of the photosynthetic system will be modeled as a QD, where the electronic structure of subunits will be described by the EHTB calculated on the basis of the 2D harmonic oscillator 

\begin{equation}\label{eq:2D-wavefunction}
\chi_{n_{x} n_{y}}\left(x,y\right)=A_{n_{x} n_{y}}\exp\left[-\dfrac{m\omega}{2\hbar}\left(x^{2}+y^{2}\right)\right]H_{n_{x}}\left(\alpha x\right) H_{n_{y}}\left(\alpha y\right),
\end{equation}

\noindent being the argument $\alpha = \sqrt{m\omega/\hbar}$, while the energy eigenvalue are written as

\begin{equation}
E_{n_{x} n_{y}}=E_{n_{x}}+E_{n_{y}}=\left(n_{x}+n_{y}+1\right)\hbar\omega,
\end{equation}

\noindent where the normalization constant is

\begin{equation}
A_{n_{x} n_{y}}=\dfrac{1}{\sqrt{2^{\left(n_{x}+n_{y}\right)}n_{x}!n_{y}!}}\sqrt{\dfrac{m\omega}{\pi\hbar}}.
\end{equation}

\noindent The kinetic energy matrix elements between two eigenstates given by Eq. \ref{eq:2D-wavefunction} can be written as a sum of overlap elements as

\begin{equation}
\begin{split}
T^{n_{x} n_{y}}_{n^{\prime}_{x} n^{\prime}_{y}}&=\dfrac{\hbar\omega}{4}\left\lbrace 2\left(n^{\prime}_{x}+n^{\prime}_{y}+1\right)S^{n_{x} n_{y}}_{n^{\prime}_{x} n^{\prime}_{y}}\right. \\
&\left.\qquad\quad -\left[n^{\prime}_{x}\left(n^{\prime}_{x}-1\right) \right]^{1/2} S^{n_{x} n_{y}}_{n^{\prime}_{x}-2; n^{\prime}_{y}}\right.\\
&\left.\qquad\quad\:\:-\left[n^{\prime}_{y}\left(n^{\prime}_{y}-1\right) \right]^{1/2} S^{n_{x} n_{y}}_{n^{\prime}_{x}; n^{\prime}_{y}-2}\right.\\
&\left.\qquad\quad\:\:\:\:-\left[\left(n^{\prime}_{x}+1\right)\left(n^{\prime}_{x}+2\right)\right]^{1/2} S^{n_{x} n_{y}}_{n^{\prime}_{x}+2; n^{\prime}_{y}}\right.\\
&\left.\qquad\quad\:\:\:\:\:\:\:-\left[\left(n^{\prime}_{y}+1\right)\left(n^{\prime}_{y}+2\right)\right]^{1/2} S^{n_{x} n_{y}}_{n^{\prime}_{x}; n^{\prime}_{y}+2} \right\rbrace,\\
\end{split}
\end{equation}

\noindent where $\omega$ is the characteristic frequency of the QD whose kinetic energy is calculated and $\mathbf{S}$ is the overlap matrix between dot states.

The EHTB method was used to study the two$-$dimensional array of quantum dots shown in Figure \ref{fig:6}a, this sketch of rings was taken from the AFM image of photosynthetic membrane of \textit{Rhodospirillum photometricum} present in the ref. \cite{scholes2011lessons}, where the large rings have a radius $R_{LHC_{I}} = 2.5$ nm and the small ones $R_{LHC_{II}} = 1.25$ nm. Defining the relation $R = \sqrt{2\hbar/\left(m\omega\right)}$ between the radius of the dot and the energy of the bound states, thereby the characteristic energy of the dots are $\omega_{LHC_{I}} = 24.4$ meV and $\omega_{LHC_{II}} = 97.5$ meV, as depicted at the left$-$hand side of Figure \ref{fig:6}b, together with the parabolic potentials that generate the basis set $\chi_{n_{x} n_{y}}(\mathbf{r})$. On the right$-$hand side, the red curve describes the EHTB effective electronic potential, $V_{eff}(\mathbf{r})$, obtained for two neighboring QDs of different sizes and the truncated basis set $\lbrace\left(n_{x}n_{y}\right)\rbrace=\lbrace\left(00\right),\left(10\right),\left(01\right)\rbrace$ which correspond to $\lbrace s,p_{x},p_{y}\rbrace$ atomic orbitals. The two main differences between the parabolic potentials (left) and the effective potential (right) can be discerned as: the two QDs can be resolved by $V_{eff}(\mathbf{r})$ but not by the parabolic potentials and, besides that, the effective potential is shallower. The EHTB effective potential is, thus, calculated for the entire array with the result shown by the surface map in Figure \ref{fig:6}c, which describes quite well the geometry of the 2D$-$QD array pictured in Figure \ref{fig:6}a.Therefore, it is concluded that the EHTB method faithfully describes the main characteristics of the system. The use of a bigger basis set would merge the effective potential of the individual dots and mixe up the array of quantum dots.

%--------------------------------------------------------------------------------
\section{CONCLUSIONS}\label{sec:conclusions}
%--------------------------------------------------------------------------------

In this work it was investigated the electronic potentials that are obtained from parametrized tight$-$binding hamiltonians. From the fundamental viewpoint, it is desirable to guarantee that the underlying physics is preserved for a given choice of tight$-$binding parameters because several paramter sets can produce similar results, whereas, from the pragmatic viewpoint, it can also be said that the non$-$uniqueness of parametrization can be used to make numerical procedures as convenient as possible.

In this work the formalim was applied to the extended H\"{u}ckel tight$-$binding (EHTB) hamiltonian, which consists of a two$-$center Slater$-$Koster method that makes explicit use of the overlap matrix. The functional form of the effective electronic potential is determined by both the tight$-$binding parameters and the basis set states. In the case of molecular systems described by Slater$-$type orbitals the ensuing potential exhibits minima at the bond regions, as a result of the resonace integral (off$-$diagonal) matrix elements. The effective electronic potential $V_{eff}(\mathbf{r})$ tends to be shallower on the atomic sites. The effect is more pronounced as the principal quantum number of the STO basis set increases because only the valence orbitals are taken into account in the EHTB method. The effect is analogous to that produced by the use of pseudo$-$potentials in first$-$principle formalisms to avoid the ion potential on the core region. The electronic potential gained from the EHTB method also describes the effects produced by the different electronegativities of the atoms, giving rise to deeper bonding potentials for the more electronegative species. The method is also applied for the 2$-$(2’$-$hydroxyphenyl)benzothiazole (HBT) molecule, after optimization of the EHTB parameters to fit \textit{ab$-$initio} calculations. The 2D potential profile is used to determine the polarity of the molecule.

As another application to the extended H\"{u}ckel theory, that has been applied extensively to atomistic systems both in chemistry and physics, it was demonstrated a procedure to model arrays of quantum dots with the EHTB method in the basis of harmonic oscillator eigenstates. The approach is versatile and describes well the characteristics of the system.

%--------------------------------------------------------------------------------
\section{ACKNOWLEDGMENTS}
%--------------------------------------------------------------------------------
The author acknowledge financial support from Brazilian agencies FAPERJ, INCT $-$ Carbon Nanomaterials, INCT $-$ Materials Informatics and INCT $-$ INEO for financial support. G.C. gratefully acknowledge FAPERJ, grant number E$-$26$/$200.627$/$2022 and E$-$26$/$210.391/2022 (project Jovem Pesquisador Fluminense process number 271814) for financial support. The authors also acknowledge the computational support of N\'{u}cleo Avan\c{c}ado de Computa\c{c}\~ao de Alto Desempenho (NACAD/COPPE/UFRJ) and Sistema Nacional de Processamento de Alto Desempenho (SINAPAD).

 \bibliographystyle{elsarticle-num} 
 \bibliography{UEPE-bib}

\end{document}